\documentclass[11pt]{amsart}

\usepackage{amsmath, amssymb}
\usepackage{mathrsfs}
\usepackage{hyperref}
\usepackage[margin=3cm]{geometry}

\newcommand{\calC}{\mathcal{C}}

\newcommand{\calE}{\mathcal{E}}

\newcommand{\calI}{\mathcal{I}}

\newcommand{\bfR}{\mathbb{R}}

\newcommand{\lie}{\mathscr{L}}

\newcommand{\dual}[1]{\left<{#1}\right>}
\newcommand{\im}{\mathrm{Im}\,}

\newcommand{\contraction}{\vrule height 0pt depth 0.4pt width 3pt
  \vrule height 7pt depth 0.4pt \kern 3pt}

\newcommand{\be}{\begin{equation}}
\newcommand{\ee}{\end{equation}}

\newtheorem{definition}{Definition}
\newtheorem{theorem}[definition]{Theorem}
\newtheorem{prop}[definition]{Proposition}
\newtheorem{lemma}[definition]{Lemma}
\newenvironment{mproof}{\textbf{Proof:}\,}{\hfill$\Box$}
\newenvironment{remark}{\textbf{Remark:}\,}{}

\newcommand{\arxiv}[1]{\href{http://www.arxiv.org/#1}{\textit{#1}}}

\newcommand{\proj}{\mathbf{h}}
\newcommand{\gotg}{\mathfrak{g}}
\newcommand{\JC}{J^{\mathrm{n.h.}}}

\newcommand{\oxi}{{\overline{\xi}}}
\newcommand{\txi}{{\tilde{\xi}}}
\newcommand{\hxi}{{\hat{\xi}}}

\newcommand{\tJ}{\tilde{J}}
\newcommand{\tJC}{\tilde{J}^{\mathrm{n.h.}}}

\newcommand{\vfT}{\mathbf{T}}

\newcommand{\tF}{{\tilde{F}}}
\newcommand{\tL}{{\tilde{L}}}
\newcommand{\tV}{{\tilde{V}}}
\newcommand{\tX}{{\tilde{X}}}
\newcommand{\tY}{{\tilde{Y}}}
\newcommand{\tZ}{{\tilde{Z}}}

\newcommand{\tcalC}{{\tilde{\calC}}}

\title{The momentum map for nonholonomic field
  theories with symmetry}
\author{Joris Vankerschaver}

\address{Joris Vankerschaver: 
Department of Mathematical Physics and Astronomy,
Ghent University, Krijgslaan 281,
B-9000 Ghent, Belgium}
\email{Joris.Vankerschaver@UGent.be}

\thanks{The author is a Research Assistant of the Research Foundation
  --- Flanders (FWO-Vlaanderen).}
\date{}

\begin{document}

\begin{abstract}
  We introduce a suitable generalization of the momentum map for
  nonholonomic field theories and prove a covariant form of the
  nonholonomic momentum equation.  We show that these covariant
  objects coincide with their counterparts in mechanics by making
  the transition to the Cauchy formalism.
\end{abstract}

\maketitle

\section{Introduction}  

In this note, we study nonholonomic field theories with symmetry.  Our
goal is to show that the results obtained in the context of mechanical
systems, such as the nonholonomic momentum map and the associated
Noether theorem, have a natural counterpart in covariant field theory.
We will mainly be concerned with the so-called \emph{multisymplectic}
approach to field theories (see \cite{fields, gimmsyI, saunders} and
the references therein).  

In section~\ref{sec:lagrangian} we review the multisymplectic
treatment of first-order Lagrangian field theories, with special
emphasis, in subsection~\ref{sec:nonhol}, on the inclusion of
nonholonomic constraints into this picture.  The rest of the paper is
then devoted to studying the action of a symmetry group: as a
warming-up, we treat in section~\ref{sec:symmfree} the case where no
constraints are present.  We review the covariant Noether theorem in a
way suitable for generalization to the constrained case.  In
section~\ref{sec:symmconstr}, we introduce constraints into the
framework and we study the implications for the Noether theorem.
Finally, in section~\ref{sec:cauchy} we break covariance to make the
link with the geometric structures known from nonholonomic mechanical
systems with symmetry.

\section{Lagrangian first-order field theories} \label{sec:lagrangian} 

\subsection{Notations}

Let $\pi: Y \rightarrow X$ be a fibre bundle of rank $m$, with $(n +
1)$-dimensional orientable base space $X$.  In addition, we will fix a
volume form $\mu$ on $X$.  Typically, $X$ will represent space-time
and the sections of $\pi$ will be the field configurations that we
wish to study.  For example, in electromagnetism, $Y$ is the cotangent
bundle $T^\ast X$ and the fields are $1$-forms representing the
electromagnetic potential.  For other physically relevant examples, we
refer to \cite{gimmsyI}.  

From time to time, it will be handy to consider coordinate expressions
of the objects involved: to this end, we choose a coordinate system
$(x^1, \ldots, x^{n + 1})$ on $X$ such that $\mu$ is locally given by
$\mu := d^{n + 1} x = dx^1 \wedge \cdots \wedge dx^{n + 1}$.  On $Y$,
we will choose a coordinate system $(x^\mu, y^a)$ adapted to the
projection $\pi$ (where $a = 1, \ldots, m$). On the first jet bundle
$J^1\pi$ we then have the induced coordinate system $(x^\mu, y^a,
y^a_\mu)$.  We will denote the projection of $J^1\pi$ onto $Y$ by
$\pi_{1, 0}$, and that onto $X$ by $\pi_1$ (such that $\pi_1 = \pi
\circ \pi_{1, 0}$).  The bundle of $\pi_1$-vertical (resp. $\pi_{1,
  0}$-vertical) vectors on $J^1\pi$ will be denoted by $V\pi_1$
(resp. $V\pi_{1, 0}$). 

For later use we also mention here a particular vector-valued $(n +
1)$-form $S_\mu$ on $J^1\pi$, called the vertical endomorphism (see
\cite{saunders}).  In coordinates, $S_\mu$ reads 
\[
  S_\mu = (dy^a - y^a_\nu d x^\nu) \wedge d^n x_\mu \otimes
  \frac{\partial}{\partial y^a_\mu},  
\]
where $d^nx_\mu$ is a short-hand notation for
$\frac{\partial}{\partial x^\mu} \contraction d^{n + 1}x$.  

\subsection{The Cartan form}

Given a regular first-order Lagrangian $L$, one can construct the
associated Cartan $(n + 1)$-form $\Theta_L$ and the multisymplectic
form $\Omega_L = -d\Theta_L$.  The coordinate expression of $\Theta_L$
is given by
\[
  \Theta_L = \frac{\partial L}{\partial y^a_\mu} (dy^a - y^a_\nu
  dx^\nu) \wedge d^nx_\mu + L d^{n + 1}x. 
\]
We will not dwell into the precise intrinsic definition of these
objects any further, but instead we refer the reader to \cite{fields,
  gimmsyI, saunders} and the references therein.

In this note, we will mainly consider the so-called De Donder-Weyl
equation (see \cite{saunders}): a connection $\Upsilon$ on $\pi_1$
with horizontal projector $\proj$ is said to be a solution of the De
Donder-Weyl equation if
\be \label{eqDDW}
  i_\proj \Omega_L = n \Omega_L. 
\ee

If $\proj$ is a solution of (\ref{eqDDW}) and $L$ a regular
Lagrangian, then a section $\psi$ of $\pi_1$ is an integral section of
$\proj$ if $\psi = j^1\phi$ for a section $\phi$ of $\pi$ (implying
that $\Upsilon$ is semi-holonomic) and, in addition, $j^1\phi$
satisfies the Euler-Lagrange equations: 
\be \label{eqEL}
 \frac{d}{d x^\mu} \left( \frac{\partial L}{\partial
    y^a_\mu}(j^1\phi) \right) - \frac{\partial L}{\partial
  y^a}(j^1\phi) = 0.  
\ee 
See \cite{nhfields02} for a proof of this statement.

\subsection{Nonholonomic constraints} \label{sec:nonhol}

In this section, we will briefly show how to derive the nonholonomic
equations of motion for a system with Lagrangian $L$ and a set of
constraints represented by a submanifold $\calC$.  For a more detailed
treatment, we refer to \cite{nhfields02, p2005}.  

Let $\calC$ be a $k$-codimensional submanifold of $J^1\pi$, with
$\pi_{1, 0}(\calC) = Y$ and such that
$(\pi_{1, 0})_{|\calC}: \calC \rightarrow Y$ is a subbundle of
$\pi_{1, 0}$.  The submanifold $\calC$ will represent some external
(nonholonomic) constraints imposed on the system.  Assume that $\calC$
is locally given by the vanishing of $k$ independent functions
$\varphi^\alpha$ and consider the subset $F$ of $\wedge^{n +
  1}(T^\ast J^1\pi)$ spanned by $\Phi^\alpha =
S^\ast_\mu(d\varphi^\alpha)$, where $S_\mu$ is the vertical
endomorphism on $J^1\pi$.  In coordinates, we
have
\[
  \Phi^\alpha = \frac{\partial \varphi^\alpha}{\partial y^a_\mu} (dy^a
  - y^a_\nu d x^\nu) \wedge d^n x_\mu. 
\]
The $(n + 1)$-forms $\Phi^\alpha$ are linearly independent because of
the initial assumption that $(\pi_{1,0})_{|\calC}$ is a subbundle of
$\pi_{1, 0}$.  Hence, $F$ is a subbundle of $\wedge^{n + 1}(T^\ast
J^1\pi)$.

In the presence of nonholonomic constraints, the field
equations become
\be \label{constrEL}
  \frac{d}{d x^\mu} \left( \frac{\partial L}{\partial
    y^a_\mu}(j^1\phi) \right) - \frac{\partial L}{\partial y^a} =
   \lambda_{\alpha\mu} \frac{\partial \varphi^\alpha}{\partial
     y^a_\mu},
\ee
together with the constraint that $j^1\phi \in \calC$ (see
\cite{nhfields02}).  Accordingly, the unconstrained De Donder-Weyl
equations are replaced by the following conditions along $\calC$:
\be \label{constrDDW}
  i_\proj \Omega_L - n \Omega_L \in \calI(F) \quad \text{and $\im
    \proj \subset T\calC$},
\ee
where $\calI(F)$ is the ideal generated by $F$.  The terms on the
right-hand side of (\ref{constrEL}) and (\ref{constrDDW}) represent
the constraint forces that keep the section $j^1\phi$ constrained to
$\calC$.  The unknown multipliers $\lambda_{\alpha\mu}$ should be
determined from the condition that $j^1\phi \in \calC$. 

\begin{remark}
In general, the constraints represented by the submanifold $\calC$ are
nonlinear.  Linear constraints can be treated as a special case of
this formalism by considering a distribution $D$ on $Y$ and taking
$\calC$ to be
\[
  \calC = \left\{ j^1_x \phi \in J^1\pi: \im T_x\phi \subset D_{\phi(x)}
  \right\}. 
\]
If $D$ is annihilated by the $k$ one-forms $A^\alpha_a dy^a +
B^\alpha_\mu dx^\mu$, then $\calC$ is given by the vanishing of the
$kn$ functions $\varphi^\alpha_\mu = A^\alpha_a y^a_\mu +
B^\alpha_\mu$.  Whenever $D$ is integrable, these constraint functions
can be written as total derivatives with respect to $x^\mu$ of
functions on $Y$, in which case the constraints can reasonably be said
to be holonomic.  This case is treated in far greater detail in
\cite{krupkova}. 
\end{remark}

\subsection{Connections on $\pi_1$}

In this section, we will prove a number of straightforward properties
of connections on $\pi_1$ that will be useful later on.  The reader is
referred to \cite{saunders} for a more comprehensive treatment.

We recall that a connection $\Upsilon$ on $\pi_1$ is said to be
\emph{semi-holonomic} if the associated horizontal projector $\proj$
satisfies $i_\proj \theta = 0$ for each contact one-form $\theta$.  In
coordinates, if
\[ 
  \proj = dx^\mu \otimes \left( \frac{\partial}{\partial x^\mu} +
  \Gamma^a_\mu \frac{\partial}{\partial y^a} + \Gamma^a_{\mu\nu}
  \frac{\partial}{\partial y^a_\nu} \right),
\]
semi-holonomicity implies that $\Gamma^a_\mu = y^a_\mu$.  This implies
that any integral section of $\proj$ is automatically the prolongation
of a section of $\pi$. 

\begin{lemma} \label{lemma:secondorder}
  For each semi-holonomic connection $\Upsilon$ with horizontal
  projector $\proj$, the following holds:
  \[
    i_\proj \Theta_L = n \Theta_L + L \mu. 
  \]
\end{lemma}
\begin{mproof}  
We give the proof in coordinates.  For any connection $\proj$, we have
\[
  i_\proj d^{n + 1} x = (n + 1) d^{n + 1} x \quad \text{and} \quad
  i_\proj d^n x_\mu = n d^n x_\mu. 
\]
Therefore, 
\[
  i_\proj \Theta_L = \frac{\partial L}{\partial y^a_\nu} i_\proj
  \theta^a \wedge d^nx_\nu + n\frac{\partial L}{\partial y^a_\nu}
  \theta^a \wedge d^nx_\nu + (n + 1) L d^{n + 1} x,
\]
where we have introduced the contact forms $\theta^a = dy^a - y^a_\mu
dx^\mu$.  If $\proj$ is semi-holonomic, the first term on the
right-hand side is zero and we obtain the desired expression. 
\end{mproof}

This lemma can be seen as the jet-bundle analogue of the well-known
fact in Lagrangian mechanics that $i_X \theta_L = \Delta(L)$ for any
second-order vector field $X$, where $\theta_L$ is the Cartan one-form
corresponding to $L$, and $\Delta$ the Liouville vector field.

\begin{lemma} \label{lemma:imvert}
  Let $X$ be a vertical vector field on $Y$ and $X^{(1)}$ its
  prolongation to $J^1\pi$. If $\Upsilon$ is a semi-holonomic
  connection on $\pi_1$ with horizontal projector $\proj$, then the
  Fr\"olicher-Nijenhuis bracket $[X^{(1)}, \proj]$ is a 
  vector-valued one-form taking values in $V\pi_{1, 0}$. 
\end{lemma}
\begin{mproof}
If $X = X^a \frac{\partial}{\partial
  y^a}$, then
\[
  X^{(1)} = X^a \frac{\partial}{\partial y^a} + \left(
  \frac{\partial X^a}{\partial x^\mu} + \frac{\partial X^a}{\partial
    y^b} y^b_\mu \right) \frac{\partial}{\partial y^a_\mu}. 
\]
(see e.g. \cite{saunders}) For the bracket, we have that $[X^{(1)},
  \proj] = \lie_{X^{(1)}} \proj$ and a straightforward calculation
then shows that this is a semi-basic vector-valued one-form taking
values in $V\pi_1$.  We now focus on the coefficient of $dx^\mu
\otimes \frac{\partial}{\partial y^a}$, which is just
\[
  X^{(1)}(\Gamma^a_\mu) - \left( \frac{\partial X^a}{\partial x^\mu} +
  \Gamma^b_\mu \frac{\partial X^a}{\partial y^b} \right). 
\]
This coefficient is easily seen to vanish when $\Gamma^a_\mu =
y^a_\mu$, i.e. when $\proj$ is semi-holonomic, which completes the
proof. 
\end{mproof}

As a corollary, we note that this lemma implies that the contraction
of $[X^{(1)}, \proj]$ with a semi-basic form (in particular with
$\Theta_L$) vanishes.

\section{Symmetry in the absence of nonholonomic
  constraints} \label{sec:symmfree} 

Let $G$ be a Lie group acting on $Y$ by bundle automorphisms $\Phi_g$
over the identity in $X$.  The assumption that $G$ acts vertically is
probably superfluous, but for the sake of clarity we will assume it
nevertheless.  

The Lie group $G$ acts on $J^1\pi$ by prolonged bundle automorphisms,
i.e. $j^1\Phi_g ( j^1_x \phi ) = j^1_x( \Phi_g \circ \phi )$.  Now,
let $L \in C^\infty(J^1\pi)$ be a $G$-invariant Lagrangian.  The action
of $G$ on $J^1\pi$ is called \emph{Lagrangian} if, for each $\xi \in
\gotg$, there exists an $n$-form $J_\xi$ (depending linearly on $\xi$)
such that $i_{\xi_{J^1\pi}} \Omega_L = d J_\xi$, where $\xi_{J^1\pi}$
denotes the infinitesimal generator corresponding to $\xi$.  In this
case, the map $J: J^1\pi \rightarrow \wedge^n(T^\ast J^1\pi) \otimes
\gotg^\ast$ defined by $\dual{J, \xi} := J_\xi$ is called the
\emph{covariant momentum map} for the action $\Phi$.

In general, we can also consider actions of $G$ on $J^1\pi$ that are
not prolonged actions of an action on $Y$, but in this note we will
nevertheless restrict ourselves to this special case.  It is easy to
see that Lagrangian actions satisfy $\lie_{\xi_{J^1\pi}} \Omega_L =
0$; for prolonged actions we have in addition that
$\lie_{\xi_{J^1\pi}} \Theta_L = 0$ (see lemma~\ref{lemma:action}).

If $G$ acts on $J^1\pi$ by prolonged bundle automorphisms, then for
each $\xi \in \gotg$ the infinitesimal generator $\xi_{J^1\pi}$ on
$J^1\pi$ is the prolongation of the infinitesimal generator $\xi_{Y}$
on $Y$.  From now on, we will denote $\xi_{J^1\pi}$ by $\xi^{(1)}$. 

\begin{lemma} \label{lemma:action}
  The Cartan $(n + 1)$-form $\Theta_L$ is invariant 
  with respect to the action of $G$ lifted to $J^1\pi$:
  \[
    \lie_{\xi^{(1)}} \Theta_L = 0. 
  \]
\end{lemma}
\begin{mproof} See \cite[p. 45]{gimmsyI}. \end{mproof}

For a prolonged action, there always exists a covariant
momentum map which is explicitly given by
\[
  J_\xi = i_{\xi^{(1)}} \Theta_L. 
\]
(see \cite[p. 45]{gimmsyI}).  The importance of the covariant momentum
map lies in the \emph{covariant Noether theorem}, first proved in
\cite{gimmsyI}.  

\begin{prop}[Covariant Noether theorem]  \label{prop:noether} Let
  $\Upsilon$ be a connection on $\pi_1$ such that the associated
  horizontal projector $\proj$ is a solution of the unconstrained De
  Donder-Weyl equation (\ref{eqDDW}).  For every $\xi \in \gotg$, the
  momentum map $J_\xi$ is constant on integral sections of $\proj$:
  \[
     d_\proj J_\xi = 0. 
  \]
\end{prop}
\begin{mproof}
In this proof, as well as in the remainder of this note, we make
frequent use of some elementary properties of the
Fr\"olicher-Nijenhuis bracket.  For the sake of completeness, we have
summarized these properties in the appendix.

We have 
\begin{align}
  d_\proj J_\xi & =  d_\proj i_{\xi^{(1)}} \Theta_L \nonumber \\
    & =  (i_\proj d - d i_\proj) i_{\xi^{(1)}} \Theta_L\nonumber \\
    & =  i_\proj \lie_{\xi^{(1)}} \Theta_L - i_\proj i_{\xi^{(1)}}
    d\Theta_L - di_\proj i_{\xi^{(1)}} \Theta_L.  \label{eqJ} 
\end{align}

In the last expression, the first term vanishes because of
lemma~\ref{lemma:action}.  The second term can be rewritten by using
the field equations (note that $\proj(\xi^{(1)}) = 0$ as $\xi^{(1)}$
is $\pi_1$-vertical):
\[
  i_\proj i_{\xi^{(1)}} d\Theta_L =
  i_{\xi^{(1)}} i_\proj d\Theta_L = - n i_{\xi^{(1)}} \Omega_L, 
\]
whereas for the last term we have, using lemma~\ref{lemma:imvert}, 
\begin{align*}
  d i_\proj i_{\xi^{(1)}} \Theta_L & = d i_{\xi^{(1)}}
  i_\proj \Theta_L \\
    & = d i_{\xi^{(1)}}\left( n\Theta_L + L \mu \right).
\end{align*}
Now, $i_{\xi^{(1)}} (L \mu) = 0$ and so we obtain 
\[
  d_\proj J_\xi = n i_{\xi^{(1)}} \Omega_L - nd
  i_{\xi^{(1)}} \Theta_L = - n \lie_{\xi^{(1)}} \Theta_L = 0,  
\]
again due to the invariance of $\Theta_L$.
\end{mproof}

\begin{remark}
  In \cite[p. 45]{gimmsyI}, the authors prove a slightly different
  Noether theorem.  They show that, if $\phi$ is a solution of the
  field equations, then $d (j^1\phi)^\ast J_\xi = 0$. It is not hard
  to prove that, for any $k$-form $\alpha$ on $J^1\pi$,
  $(j^1\phi)^\ast d_\proj \alpha = d (j^1\phi)^\ast \alpha$ if and
  only if $j^1\phi$ is an integral section of $\proj$.
  Proposition~\ref{prop:noether} therefore implies that $d
  (j^1\phi)^\ast J_\xi = 0$.  The proof of the Noether theorem in
  \cite{gimmsyI} is more straightforward; our proof has the advantage
  that it will be easily extendible to the case where nonholonomic
  constraints are present.
\end{remark}

\section{The constrained momentum map} \label{sec:symmconstr}

In this section, we study the case of a constrained field theory, with
regular Lagrangian $L$ and constraint submanifold $\calC$ satisfying
the assumptions of section~\ref{sec:nonhol}.  The constrained De
Donder-Weyl equations are then given by (\ref{constrDDW}).

Suppose now that in addition to these nonholonomic constraints, there
is also a symmetry group $G$ acting on $J^1\pi$ by prolonged bundle
automorphisms, such that both the Lagrangian $L$ and the constraint 
manifold $\calC$ are $G$-invariant, i.e.
\[
  L \circ j^1\Phi_g = L \quad \text{as well as} \quad 
    j^1\Phi_g(\calC) \subset \calC
\]
for all $g \in G$. In general, as in the case of nonholonomic
mechanics (see \cite{Sn, bloch96, frans}), it will no longer be true
that these symmetries give rise to conserved quantities; the precise
link will be made clear by the \emph{nonholonomic momentum equation}
or constrained Noether theorem (theorem~\ref{th:eqmom}).  Our
treatment extends the one in \cite{frans}; we refer to that paper, as
well as to \cite{Sn, bloch96} and the references therein, for more
information about the nonholonomic momentum equation in mechanics.

We first introduce the following distribution:
\[
  \calE(\gamma) = \{ v \in T_\gamma J^1\pi: i_v(S_\mu^\ast
  d\varphi_\alpha) = 0 \text{ for each $\alpha = 1, \ldots, k$}\}
  \quad \text{where $\gamma \in \calC$}.
\] 
For a given $\gamma \in \calC$ we consider all elements
$\xi$ of the Lie algebra $\gotg$ such that $\xi^{(1)}(\gamma) \in
\calE(\gamma)$.  The set of all such $\xi$ we denote by
$\gotg^\gamma$.  We take $\gotg^\calE$ to be the disjoint union of all
these spaces $\gotg^\gamma$ and we assume that $\gotg^\calE$ can be
given the structure of a bundle over $\calC$.

With these elements in mind, we define the \emph{constrained momentum
  map} as a map $\JC: \calC \rightarrow \wedge^n (J^1\pi) \otimes
g^\calE$, constructed as follows.  With every section $\oxi$ of
$\gotg^\calE$, one may associate a vector field $\txi$ on $J^1\pi$ by
putting $\txi(\gamma) = (\oxi(\gamma))_{J^1\pi}(\gamma)$.  Remark that
$\txi$ is a section of $\calE$.  We then define $\JC_\oxi$ along
$\calC$ as
\[
    \JC_\oxi = i_\txi \Theta_L. 
\] 

The importance of the nonholonomic momentum map lies in the
nonholonomic momentum equation:
\begin{theorem}[Nonholonomic momentum equation] \label{th:eqmom}
  Let $\Upsilon$ be a connection on $\pi_1$ such that the associated
  horizontal projector $\proj$ is a solution of the constrained De
  Donder-Weyl equation.  Assume furthermore that $G$ is a Lie group
  acting on $J^1\pi$ in the way described above.  Then the
  nonholonomic momentum map satisfies the following equation:
  \[
    d_\proj \JC_\oxi = \lie_\txi(L \mu) \quad \text{along $\calC$}. 
  \]
\end{theorem}
\begin{mproof}
Equation (\ref{eqJ}) from the proof of proposition~\ref{prop:noether}
can be used without modification:
\begin{align*}
    d_\proj \JC_\oxi & =  i_\proj \lie_\txi \Theta_L - i_\proj i_\txi
    d\Theta_L - di_\proj i_\txi \Theta_L \\
      & =  i_\proj \lie_\txi \Theta_L + i_\txi(n \Omega_L + \zeta)
    - n \lie_\txi \Theta_L + ni_\txi d\Theta_L. 
\end{align*}
In this expression, we have substituted the constrained De Donder-Weyl
equation: $\zeta$ is an element of $\calI(F)$.  As $\zeta$ can be
written as $\zeta = \lambda_{\alpha\mu} dx^\mu \wedge f^\alpha$
(see \cite{p2005}), with $f^\alpha$ taking values in the bundle 
$F$, we may conclude that $i_\txi \zeta = 0$.  Therefore, we end up
with
\begin{align*}
    d_\proj \JC_\txi & = i_\proj \lie_\txi \Theta_L - n\lie_\txi
      \Theta_L \\
      & = \lie_\txi i_\proj \Theta_L - i_{[\txi, \proj]} \Theta_L -
      n \lie_\txi \Theta_L \\
      & = \lie_\txi(L\mu),
\end{align*}
where we have used the remark following lemma \ref{lemma:imvert} to
conclude that $i_{[\txi, \proj]} \Theta_L = 0$.
\end{mproof}

We finish by noting that in the case where $\txi$ can be written as
$\xi^{(1)}$ (for example, when $\oxi$ is a constant section), we may
conclude from the $G$-invariance of $L$ that $d_\proj \JC_\txi = 0$.
In general, though, this will not be the case.

\section{The Cauchy formalism} \label{sec:cauchy}

Up until now, all of our results have been derived in a purely
covariant setting where all of the coordinates on the base space $X$
are treated on an equal footing.  In particular, there is no
distinguised time coordinate.  We will now assume that the
Euler-Lagrange equations associated to the Lagrangian $L$ describe an
(hyperbolic) initial-value problem.  In this case, it is meaningful to
single out a global direction of time and break covariance by making
the transition to the space of Cauchy data.  We can then rephrase the
field equations accordingly as a time-dependent mechanical system on
an infinite-dimensional configuration space (see \cite{symm04,
  gimmsyII, santamaria}).

This is done by fixing a particular diffeomorphism $\Psi: \bfR \times
M \rightarrow X$, where $M$ is an $n$-dimensional manifold (and where
we tacitly assume that the topology of $X$ is such that $\Psi$ can
indeed be globally defined), thus singling out a ``splitting'' of $X$
into space and time.  To avoid the technical matters arising when
considering the behaviour of the field ``at infinity'', we assume that
$M$ is compact.  We define the space $\tX$ to consist of all
embeddings $\tau$ of $M$ into $X$ such that there exists a $t \in
\bfR$ for which $\tau = \Psi(t, \cdot)$.  Hence, there is a one-to-one
correspondence between $\bfR$ and $\tX$.  This correspondence, or the
existence of the diffeomorphism $\Psi$, induces a distinguished vector
field $\vfT$ on $X$, defined at $x \in X$, by
\[
  \vfT(x) = \frac{d}{ds} \Psi(s, u) \Big|_{s = t}, \quad \text{where
    $x = \Psi(t, u)$}. 
\]
For the sake of convenience, we will assume that $M$ is equipped with
a volume form $\mu_M$ such that $\mu := dt \wedge \mu_M$ is a
volume form for $X$, where $t$ is a global coordinate labelling $\tX$.

We define the \emph{space of Cauchy data} (denoted by $\tZ$) as the
space of embeddings $\kappa: M \hookrightarrow J^1\pi$ for which there
exists a section $\phi$ of $\pi$ and an element $\tau$ of $\tX$ such
that $\kappa = j^1\phi \circ \tau$.  For more information on this
space (which can be given the structure of a smooth manifold in some
suitable sense), we refer the reader to \cite{bsf, gimmsyII, michor,
  santamaria}.  There exists a convenient way of viewing the tangent
bundle of $\tZ$: a tangent vector $v \in T_\kappa \tZ$ can be seen as
a section of $\Gamma(\kappa^* TJ^1\pi)$ (a vector field along
$\kappa$).  There exist similar interpretations of $T\tX$ and $T\tY$.

A vector field $V$ on $J^1\pi$ induces a vector field $\tV$ on
$\tZ$ by composition: $\tV(\kappa) = V \circ \kappa$. 
Similarly, an $(n + k)$-form $\alpha$ on $J^1\pi$ induces a $k$-form
$\tilde{\alpha}$ on $\tZ$ by integration:
\be \label{induction}
  \tilde{\alpha}(\kappa)(\tV_1, \ldots, \tV_k) = 
    \int_M \kappa^\ast i_{\tV_1 \wedge \cdots \wedge \tV_k} \alpha. 
\ee
By use of this correspondence, the multisymplectic form $\Omega_L$ and
the volume form $\mu$ induce respectively a two-form $\tilde{\Omega}_L$ and
a one-form $\tilde{\mu}$ on $\tZ$, whereas the Lagrangian $L$ can be
seen to induce a function on $\tZ$:
\[
  \tL(\kappa) = \int_M \kappa^\ast i_\vfT (L \mu). 
\]
Strictly speaking, on the right-hand side of this expression one
should replace $\vfT$ by an arbitrary vector field $V$ on $J^1\pi$
projecting down to $\vfT$, but since $L \mu$ is semi-basic, the
contraction does not depend on $V$ but only on $\vfT$.

\begin{remark}
It has been shown that the covariant field equations induce a
dynamical system $\Gamma$ on $\tZ$ whose determining equations are
formally identical to those of a time-dependent mechanical system with
an infinite-dimensional configuration space (see \cite{gimmsyII, santamaria}):
\be \label{EoMfree}
  i_\Gamma \tilde{\Omega}_L = 0 \quad \text{and} \quad i_\Gamma
  \tilde{\mu} = 1. 
\ee
In \cite{p2005}, we showed that in the case of nonholonomic field
theory, the induced dynamical system on $\tZ$ is determined by
\be \label{EoMconstrained}
  i_\Gamma \tilde{\Omega}_L\big|_{\tcalC} \in \tF 
     \quad \text{and} \quad \Gamma \in T\tcalC, 
\ee
where $\tF$ is a codistribution induced by $F$ and $\tcalC$ is the
subset of $\tZ$ induced by $\calC$ and defined as 
\[
   \tcalC = \{ \kappa \in \tZ : \im \kappa \subset \calC \}.
\]
In both the constrained and the unconstrained case, a connection
$\Upsilon$ solving the covariant field equations induces a vector
field $\Gamma$ on $\tZ$ which is a solution of the corresponding
dynamical system on $\tZ$.  In the unconstrained case, this dynamical
system is given by (\ref{EoMfree}), whereas in the constrained case
the equations of motion are given by (\ref{EoMconstrained}).  The
precise relation between $\proj$ and $\Gamma$ is
\be \label{solution}
  \Gamma(\kappa) = \proj( T j^1\phi(\vfT) ) \circ \kappa,
\ee
where we have decomposed $\kappa$ as $\kappa = j^1\phi \circ \tau$. 
With some abuse of notation, we will also write $\Gamma =
\proj(\vfT)$. 
\end{remark}

In the next sections, we will exhibit the structures on $\tZ$ induced
by the (nonholonomic) momentum map and we will show how the covariant
momentum equation give rises to a momentum equation on $\tZ$ which is
formally identical to the one encountered in nonholonomic mechanics
(see for example \cite{bloch96, frans}). 

By (\ref{induction}), the component $J_\xi: J^1\pi \rightarrow
\wedge^n(J^1\pi)$ of the covariant momentum map induces a map $\tJ_\xi
\in C^\infty(\tZ)$ on the space of Cauchy data:
\[
  \tJ_\xi(\kappa) = \int_M \kappa^\ast J_\xi. 
\]
In the constrained case, there is a similar definition for the map
$\tJC_\xi$ in the Cauchy formalism, induced by the component $\JC_\xi$
of the constrained momentum map.  Note that $\JC_\xi$ is defined along
$\calC$.  

\subsection{The unconstrained case}

We now turn to proving the analogue of Noether's theorem in the Cauchy
framework.  There are essentially two ways in which one could approach
this problem: either by directly defining the action of $G$ on $\tZ$
and using the standard techniques known from mechanics, or by showing
that the covariant Noether theorem leads in a straightforward way to
the Noether theorem on the space of Cauchy data.  We choose to follow
the second approach, as it allows us to postpone to the very end all
of the technical matters associated with the calculus on
infinite-dimensional manifolds.

\begin{prop}
  Let $\Upsilon$ be a connection in $\pi_1$ such that the associated
  horizontal projector $\proj$ is a solution of the De Donder-Weyl
  equation (\ref{eqDDW}).  Let $\tJ$ be the momentum map associated to
  the covariant momentum map $J$.  Then Noether's theorem holds:
  $\Gamma( \tJ_\xi ) = 0$ for all $\xi \in \gotg$, where $\Gamma$ is a
  solution to the equations of motion (\ref{EoMfree}) in the Cauchy
  formalism.
\end{prop}
\begin{mproof}
We will use the following characterisation of the exterior derivative
$d\tJ_\xi$ in terms of $dJ_\xi$:
\[
  \dual{\tV, d\tJ_\xi}(\kappa) = \int_M \kappa^\ast( i_\tV
  dJ_\xi ), 
\]
for an arbitrary vector field $\tV$ on $\tZ$.  For a proof, we refer
to \cite[prop.~3.3.9]{santamaria} or to the expressions used in
\cite[lemma~5.1]{gimmsyII}.

The embedding $\kappa : M \hookrightarrow J^1\pi$ can be written as
$\kappa = j^1 \phi \circ \tau$.  Without loss of generality, we may
take $\phi$ to be a solution of the field equations.  This lies at the
heart of the Cauchy analysis: $\kappa$ specifies the values of the
fields and their derivatives on a hypersurface and due to the
(supposed) hyperbolicity of the equations of motion, the subsequent
evolution is then fixed (and given by $j^1\phi$).  Formally, let $t
\mapsto c(t)$ be an integral curve of $\Gamma$ such that $c(0) =
\kappa$.  Then $j^1_x \phi = [c(t)](u)$, where $x = \Phi(t, u)$.

We then have, noting that $\proj(\vfT) = T j^1\phi( \vfT )$,
\[
  \dual{\Gamma, d\tJ_\xi}(\kappa) = \int_M \kappa^\ast( i_\Gamma d
  J_\xi) = \int_M \tau^\ast (j^1\phi)^\ast( i_{\proj(\vfT)} dJ ) =
  \int_M \tau^\ast i_\vfT ( (j^1\phi)^\ast dJ ).
\] 
As we pointed out in the remark following
proposition~\ref{prop:noether}, one can check that
$(j^1\phi)^\ast d_\proj \alpha = d (j^1\phi)^\ast \alpha$ if and only
if $j^1\phi$ is an integral section of $\proj$.  We conclude that
\be \label{transcauchy}
  \dual{\Gamma, d\tJ_\xi}(\kappa) = \int_M \tau^\ast i_\vfT(
  (j^1\phi)^\ast d_\proj J_\xi). 
\ee

As the $\xi$-component $J_\xi$ of the covariant momentum map satisfies
Noether's theorem, i.e. $d_\proj J_\xi = 0$, we have that $\Gamma( \tJ_\xi
) = 0$.  This establishes the theorem of Noether in the Cauchy framework.
\end{mproof}

\subsection{The constrained case}

Quite surprisingly, much of the material developed in the preceding
section carries over quite naturally to the constrained case.  In
particular, for the nonholonomic momentum map, equation
(\ref{transcauchy}) still holds:
\[
  \dual{\Gamma, d\tJC_\oxi}(\kappa) = \int_M \tau^\ast i_\vfT(
  (j^1\phi)^\ast d_\proj \JC_\oxi), \quad \text{for $\kappa \in
    \calC$}, 
\]
where we attribute a similar meaning to all terms involved: $\proj$ is
a solution of the constrained De Donder-Weyl equation, $j^1\phi$ is an
integral section of the corresponding connection and $\Gamma =
\proj(\vfT)$.  Note that $\Gamma$ is now a solution of
(\ref{EoMconstrained}). 

Now, if $\JC_\oxi$ satisfies the nonholonomic momentum
equation, then
\be \label{nhmom}
  \dual{\Gamma, d\tJC_\oxi}(\kappa) = \int_M \tau^\ast i_\vfT (
  (j^1\phi)^\ast \lie_\oxi(L\mu) ).
\ee
In the following proposition, we further elaborate the right-hand
side.  We recall that the vector field $\txi$ on $J^1\pi$ naturally
induces a vector field $\hxi$ on $\tZ$ by putting $\hxi(\kappa) = \txi
\circ \kappa$.

\begin{prop}
Let $\Upsilon$ be a connection on $\pi_1$ such that along the
constraint submanifold $\calC$ the associated horizontal projector
$\proj$ satisfies the constrained De Donder-Weyl equation.  Assume a
Lie group $G$ acts in the way described above and let $\tJC$ be the
momentum map associated to the covariant momentum map $\JC$.  Then
$\tJC$ satisfies the nonholonomic momentum equation: for all $\oxi \in
\gotg^\calE$,
\[
   \Gamma(\tJC_\oxi) = \hxi( \tL ) \quad \text{along $\calC$}. 
\]
\end{prop}
\begin{mproof}
We rewrite the right-hand side of (\ref{nhmom}) by performing exactly
the opposite manipulations as we did to obtain
eq.~(\ref{transcauchy}).  This leads to
\[ 
  \dual{\Gamma, d\tJC_\oxi}(\kappa) = \int_M \kappa^\ast i_{\proj(\vfT)}
    \lie_{\txi}(L\mu) = \int_M \kappa^\ast \lie_{\txi}
    (i_{\proj(\vfT)}(L\mu)) + \int_M \kappa^\ast i_{[\proj(\vfT), \txi]}(L\mu). 
\]
The last term vanishes as $L\mu$ is semi-basic and $[\proj(\vfT),
  \txi]$ is $\pi_1$-vertical ($\txi$ is $\pi_1$-vertical).  By
lemma~3.3.9 of \cite{santamaria}, we see that the first term on the
right-hand side equals
\[
  \int_M \kappa^\ast \lie_{\txi} (i_{\proj(\vfT)}(L\mu)) =
  \lie_\hxi( \tL ),
\]
and this proves the momentum equation in the Cauchy formalism.
\end{mproof}

\section*{Acknowledgements}

Financial support of the Research Foundation--Flanders
(FWO-Vlaanderen) is gratefully acknowledged.  I would also like to
thank Frans Cantrijn for useful discussions and a critical reading of
this manuscript, as well as Manuel de~Le\'on and David Mart\'{\i}n
de~Diego for many fruitful discussions and their kind hospitality
during several research visits to the CSIC (Madrid).

\section*{Appendix: elementary properties of the
  Fr\"olicher-Nijenhuis bracket}

In this section, we review some properties of the
Fr\"olicher-Nijenhuis bracket and the various derivations associated
to vector-valued forms on a manifold.  For a detailed treatment of the
Fr\"olicher-Nijenhuis bracket, we refer the reader to \cite{kms,
  saunders}.

Let $M$ be a manifold.  A \emph{vector-valued one-form $\proj$} is a
section of $TM \otimes T^\ast M$.  Associated to $\proj$ is a derivation
$i_\proj$ (of type $i_\ast$ and degree $0$), defined by
\[ \label{ider}
   (i_\proj \alpha)(v_0, \ldots, v_k) = \sum_{i = 0}^k (-1)^i \alpha(
\proj(v_i), v_0, \ldots, \widehat{v_i}, \ldots, v_k) \quad \text{for
  $\alpha \in \Omega^{k + 1}(M)$}. 
\]
We then define $d_\proj$ as $d_\proj = i_\proj \circ d - d \circ
i_\proj$; this is a derivation of type $d_\ast$ and degree $1$. 

Vector-valued forms of higher degree are defined accordingly as
sections of the tensor product $TM \otimes \wedge^k(T^\ast M)$.  A
vector-valued $k$-form $R$ can easily be seen to give rise to a
derivation $i_R$ of degree $k - 1$ (by virtue of a generalization of
eq. \ref{ider}) as well as a derivation $d_R$ of degree $k$.  A
vector-valued form of degree zero is simply a vector field, and the
associated derivations are in this case the contraction $i_X$ and the
Lie derivative $\lie_X$.

The Fr\"olicher-Nijenhuis bracket of a vector-valued $r$-form $R$ and
a vector-valued $s$-form $S$ is then defined as the unique
vector-valued $(r + s)$-form $[R, S]$ for which
\[ \label{FN}
  d_R \circ d_S - (-1)^{rs} d_S \circ d_R = d_{[R, S]}. 
\]
We have deliberately been vague about the nature of this bracket: most
of the time we will only need the bracket of a vector field $X$ with
a vector-valued one-form $\proj$ (which will be the horizontal
projector of a connection).  In this case, it is not hard to prove
that
\[
  [X, \proj] = \lie_X \proj. 
\]

The following lemma collects the properties of the
Fr\"olicher-Nijenhuis bracket that we will be needing in the body of
the text.  They can be suitably generalized and form part of a
well-investigated calculus, for which we refer to \cite{kms}.

\begin{lemma} \label{lemma:FN}
  Let $X$ be a vector field on $M$ and $\proj$ a vector-valued
  one-form.  Then, for any $k$-form $\alpha$ on $M$, the following
  holds:
  \begin{enumerate} 
    \item $i_X i_\proj \alpha = i_\proj i_X \alpha + i_{\proj(X)}
      \alpha$;
    \item $i_\proj \lie_X \alpha = \lie_X i_\proj \alpha - i_{[X,
        \proj]} \alpha$.   
  \end{enumerate}
\end{lemma}
\begin{mproof}
Let $\alpha$ be a $2$-form (the case of a $k$-form $\alpha$ is
completely similar) and $Y$ a vector field on $M$.  Then
\begin{align*}
  (i_X i_\proj \alpha)(Y) & =  
        \alpha(\proj(X), Y) - \alpha(\proj(Y), X) \\
      & =  (i_{\proj(X)} \alpha)(Y) + (i_\proj i_X \alpha)(Y),
\end{align*}
which confirms the first property. 

The second property (a special case of lemma~8.6 in \cite{kms}) can be
proved directly by noting that a derivation is completely determined
by its action on functions and one-forms.  For a function $f$ both
sides of the relation (2) vanish and for a one-form $\alpha$ we have
for the left-hand side
\begin{align*}
  (i_\proj \lie_X \alpha)(Y) & =  (\lie_X \alpha) (\proj(Y))
     =  \lie_X ( \alpha(\proj(Y)) ) - \alpha( [X, \proj(Y)] ) \\
    & =  \lie_X ( \alpha(\proj(Y)) ) - \alpha( (\lie_X \proj)(Y) ) 
          - \alpha( \proj([X, Y])). 
\end{align*}
Taking together the first and third term, we obtain $\lie_X ( i_\proj
\alpha )(Y)$, whereas the second term is just $i_{[X, \proj]}
\alpha(Y)$. 
\end{mproof}


\begin{thebibliography}{99}
\bibitem{Sn} L. Bates, J. \'Sniatycki: \textit{Nonholonomic
  reduction}.  Rep. Math. Phys. \textbf{93}, no. 1, 99--115. 

\bibitem{nhfields02} E. Binz, M. de Le\'on, D. Mart\'{\i}n de Diego,
  D. Socolescu: \textit{Nonholonomic Constraints in Classical Field
    Theories}.  Rep. Math. Phys \textbf{49} (2002), 151--166. 

\bibitem{bsf} E. Binz, J. \'Sniatycki, H. Fischer: \textit{Geometry
  of classical fields}.  North-Holland Mathematics Studies, 154. 
  North-Holland Publishing, Amsterdam, 1988. 

\bibitem{bloch96}  A. Bloch, P. Krishnaprasad, J. Marsden, R. Murray:
  \textit{Nonholonomic mechanical systems with symmetry}. 
  Arch. Rat. Mech. Anal. \textbf{136} (1996), no. 1, 21--99.  

\bibitem{frans}  F. Cantrijn, M. de Le\'on, J.C. Marrero,
  D. Mart\'{\i}n de Diego: \textit{Reduction of nonholonomic
    mechanical systems with symmetry}.  Rep. Math. Phys. \textbf{42}
  (1998), no. 1/2, 25--45. 

\bibitem{p2005} F. Cantrijn, M. de Le\'on, D. Martin de Diego, J. 
  Vankerschaver: \textit{Geometric aspects of classical field theories
    with nonholonomic constraints}.  Preprint, to appear in Rep. Math. 
    Phys., available as \arxiv{math-ph/0506010}. 

\bibitem{symm04} M. de Le\'on, D. Mart\'{\i}n de Diego,
  A. Santamar\'{\i}a-Merino: \textit{Symmetries in Field Theory}. 
  Int. J. Geom. Methods Mod. Phys. \textbf{1} (2004), no. 5,
  651--710. 

\bibitem{fields}  M. de Le\'on, M. McLean, L. Norris, A.R. Roca,
  M. Salgado: \textit{Geometric structures in field theory}. 
  Preprint, online available as \arxiv{math-ph/0208036}. 

\bibitem{gimmsyI} M.J. Gotay, J. Isenberg, J.E. Marsden:
  \textit{Momentum Maps and Classical Relativistic Fields.  Part I:
    Covariant Field Theory}.  Preprint, online available as
  \arxiv{physics/9801019}. 

\bibitem{gimmsyII} M.J. Gotay, J. Isenberg, J.E. Marsden:
  \textit{Momentum Maps and Classical Relativistic Fields.  Part II:
    Canonical Analysis of Field Theories}.  Preprint, online available
  as \arxiv{math-ph/0411032}. 

\bibitem{kms} I. Kol\'a\v{r}, P. Michor, J. Slov\'ak: \textit{Natural
  operations in differential geometry}.  Springer-Verlag, Berlin,
  1993. 

\bibitem{michor} A. Kriegl, P.W. Michor: \textit{The Convenient
    Setting of Global Analysis}. Mathematical Surveys and Monographs
  \textbf{53} -- AMS, Providence (RI), 1997. 

\bibitem{krupkova} O. Krupkov\'a: \textit{Partial differential
    equations with differential constraints}.  To appear in
  J. Diff. Eq. 

\bibitem{santamaria} A. Santamar\'{\i}a-Merino: \textit{M\'etodos
  Geom\'etricos en Teor\'{\i}as Cl\'asicas de Campos e Integraci\'on
  Num\'erica}.  PhD thesis, Universidad Carlos III de Madrid, 2005. 

\bibitem{saunders} D.J. Saunders: \textit{The Geometry of Jet
  Bundles}.  LMS Lecture Note Series, vol. 142. Cambridge University
  Press, 1989. 
\end{thebibliography}
\end{document}